\documentstyle[aps,prl,psfig,epsfig]{revtex}
\newcommand{\be}{\begin{equation}}
\newcommand{\ee}{\end{equation}}
\newcommand{\bea}{\begin{eqnarray}}
\newcommand{\eea}{\end{eqnarray}}

\tightenlines
\begin{document}
\draft

\title{Theoretical modeling of prion disease incubation}

\author{R.V. Kulkarni$^{1}$, A. Slepoy$^{2}$,  
R.R.P. Singh$^{1}$, D.L. Cox$^{1}$, D. Mobley$^{1}$, and F. P\'{a}zm\'{a}ndi$^{1}$ }
\address{$^1$Department of Physics, University of California, Davis, CA 95616
\\ $^2$MS 0316, Sandia National Laboratories,
P. O. Box 5800, Albuqeruque, NM 87185-0316}
\twocolumn[\hsize\textwidth\columnwidth\hsize\csname
@twocolumnfalse\endcsname

\date{\today}

\maketitle

\begin{abstract}

We present a theory for the laboratory and epidemiological data for incubation
times in infectious prion diseases. The central feature of our model
is that slow growth of misfolded protein-aggregates from small initial
seeds controls the `latent' or `lag' phase, whereas
aggregate-fissioning and subsequent spreading leads to an exponential
growth or doubling phase. Such a general framework can account for
many features of prion diseases including the striking reproducibility
of incubation times when high doses are inoculated into lab animals.
Furthermore, we explore the importance of aggregate morphology in
determining the statistics of the incubation time.  Broad incubation
time distributions arise for low infectious dose, while our calculated
distributions narrow to sharply defined onset times with increased
dose. We use our distributions to obtain a fit for the experimental
dose-incubation curves for distinct strains of scrapie and show how
features of the dose-incubation curve, specifically (a) the
logarithmic dose-dependence at high dose and (b) deviations from
logarithmic behavior at low dose can be explained within our model.
By fitting the experimental dose-incubation curves, we quantify our
model parameters and make testable predictions for experiments which
measure the time-course of infectivity.  We apply our model to
analysis of data from BSE epidemiology, iatrogenic CJD infections,
and vCJD infections; these data are consistent with incubation times dominated
by slow aggregation from a few seeds which are tens of nanometers in
size.  Within the model, small mammals derive shorter incubation times
from much more rapid protein attachment rates which we suggest to arise
from higher PrP$^c$ concentrations that might be metabolically
controlled.  Further, based upon our analysis 
we suggest that vCJD incubation times
are likely to be at the low end of previous estimates. 
\end{abstract}

\vskip 1cm

]

\narrowtext

\section{INTRODUCTION}

Understanding the factors which regulate the incubation times for
infectious prion diseases is important for assessing the risk of
illness after potential exposure as well as for developing treatments
which can delay disease onset. There are several striking aspects to
prion disease incubation, which are not well understood: (i) The
incubation times can run into years and decades \cite{prusiner82a}, 
and yet, at the
laboratory scale have been found to be highly reproducible. In fact,
the reproducibility of incubation times with dose has been used as an
independent measure of infectivity titre \cite{prusiner82b,prusiner99}.
(ii) There seem to be
distinct stages for the disease incubation after intracerebral inoculation: 
(a) Following rapid initial clearance, there is a `lag' phase (also termed 
`zero' phase \cite{dickinson79,kimberlin88}) during which
there is little or no infectivity, and 
(b) an exponential growth or doubling
phase, during which the infectivity increases exponentially with a 
well-defined doubling period \cite{manuelidis96,bolton98,beekes96}.
Understanding the lag-phase is clearly
important as any treatment strategy is more likely to succeed
before the exponential growth phase takes over. (iii) As the dose of
infection is increased in the laboratory, the incubation times become
sharply defined and saturate to a dose independent value, but as the
dose is reduced a broad distribution and a logarithmic dose dependence
results \cite{mclean00}. Such a broad distribution has also been found in
epidemiological studies of Bovine Spongioform Encephelopathy (BSE) in
England \cite{stekel96,anderson96}.
(iv) For infection across species, there is a `species
barrier' and the first passage takes considerably longer to incubate
than subsequent passages \cite{kimberlin77,kimberlin79}.
(v) While prion aggregation has been
observed {\it in vitro}, the aggregates are neuro-toxic but not
infectious \cite{post00}. These are the issues which motivate our study.

The purpose of this paper is to test the extent to which a purely
physico-chemical model can capture the main reproducible features of
prion disease incubation. In particular, we emphasize the importance
of the aggregate morphology in determining the statistics of
incubation times.  Our basic hypothesis is that the `lag phase' is
determined by growth of misfolded protein-aggregates from initial
small seeds (acquired through infection) to a typical `fissioning
dimension', whereas subsequent aggregate-fissioning and spreading
leads to exponential growth and the doubling phase. For a single seed,
the lag phase develops a broad but well defined distribution, which we
can calculate via a microscopic statistical model. Thus, when the
infection is very dilute, there is a broad distribution of incubation
times. At higher doses of infection, self-averaging due to independent
growth from many seeds leads to sharply defined incubation times. The
dose dependence and its saturation, as well as the ratio of lag time
to doubling time depends on the morphology of the aggregates, {\it
i.e.}, whether one has linear fibrils or compact higher-dimensional
aggregates. In this sense, details of incubation time distributions
provide an indirect means to infer early growth
morphologies. Alternative theoretical models which deal with the above
issues have also been developed in the literature. 
\cite{eigen96,nowak98,masel99,masel00,stumpf00,payne98,kellershohn01}

The extent to which such a model explains the experimental
phenomenology would help address the following questions: (i) {\it Are
the incubation times dominated by the nucleation and growth of
misfolded protein aggregates?}. 
(ii) {\it Are the two phases of prion disease incubation, the lag 
phase and the exponential growth phase, controlled by the same process
i.e. aggregation of misfolded proteins? }
(iii) Assuming that aggregate growth controls the incubation time scales, we
are led to ask: {\it What is the aggregate morphology during early
growth and how does it influence the dose incubation curves}
(iv) {\it Does current data inform us about the characteristic size of the
aggregates?}
(v) {\it What are the practical epidemiological
impacts from our model incubation time distributions?}  
Can we use the model to constrain estimates of the number of infections
of vCJD for example?

Employing statistical simulations of prion aggregation (based upon
cellular automata rules) we argue here that several features of prion
disease can be explained by exploring the statistics of the two phases
of the disease incubation. Within our model, we show that compact
two-dimensional aggregates can provide the observed broad distribution
of incubation times for dilute doses and at the same time account for
the typical large difference between lag time and doubling time.  We
present analytic calculations which provide a functional form for the
distribution that can be used in further epidemiological studies, and
we use these results to infer the dose dependence of the incubation
time. Furthermore, our analysis shows how the dose-incubation curve
can be related to experimental measurements of the time-course of
infectivity and in particular, testable predictions can be made using
our model for the dose dependence of the lag phase. Finally, we apply
our model to epidemiological data for BSE (mad cow disease),
iatrogenic CJD infections associated with {\it dura mater}
transplants, and a vCJD cluster from the United Kingdom, from which we
conclude that in all three cases the incubation time is dominated by
slow aggregation from a few small (tens of nanometer scale) starting
seeds. This size scale is comparable to small animals, but the
estimated attachment rates are slower by an order of magnitude or
more. We argue that the incubation time for vCJD is likely to be at
the low end of previous estimates, implying an infectious toll in the
hundreds rather than hundreds of thousands.  We speculate that the
differing attachment rate among species is regulated by the PrP$^c$
concentration which in turn is metabolically controlled.

We organize our paper as follows: Section II discusses a model
incubation time distribution, which illustrates how our basic picture
relates to dose incubation curves in prion diseases. Sections III-V
deal with microscopic models related to protein misfolding,
aggregation and fissioning. In Sections VI, VII, and VIII, we come
back to the dose incubation curves,
connections to epidemiological data and the disease phenomenology
and discuss them in context of our models and present our conclusions.

\section{A Model Distribution and Dose Incubation Curves}

We first illustrate the `bare bones' of our proposed picture for
prion disease incubation by using a model distribution, where
calculations can be done analytically. As mentioned in the
previous section, the lag phase corresponds to aggregation of misfolded
prions from the initial seed upto a `fissioning size'. This is a
stochastic process and correspondingly there will be a distribution of
aggregation times. Subsequent to this aggregation, we assume (based
upon experimental observations) that the number of seeds increases
exponentially with a well defined doubling time ($t_2$). This
exponential growth continues until the number of seeds reaches a
critical value which signals the onset of clinical symptoms and the 
end of the incubation period.

In this section, we assume the distribution of aggregation times
from a single seed, $P(t)$, to be given by
\begin{eqnarray}
P(t) & = & 0      ~~~~ t < t_0   \\
     & = & \frac{1}{n_{1}t_2}    ~~~~~ t_0 < t < t_0 + n_{1}t_2 \\
     & = & 0       ~~~~~~ t > t_0 + n_{1}t_2  
\end{eqnarray}
Here $t_0$ and $n_{1}$ are the parameters for the distribution.
It has a sharp onset at time $t_0$ and its width is taken to be
some multiple ($n_1$) of the doubling time $t_2$. 
When there are many seeds present, each
initial seed will start fissioning into two new seeds after an
aggregation time according to the above distribution. We will assume
that once a seed has fissioned once, it continues to fission or effectively
double every $t_2$ time steps, which is independent of the above
distribution. The microscopic basis for this assumption
will be explained in section V.
The incubation time is given by the mean time taken for a
given initial number of seeds ($D_i$)
to reach a critical number ($D_f$) following the
above processes.

The dose dependence of the the incubation time 
is calculated through the following steps:
 
1) First we calculate the mean `first arrival' time i.e. the mean time 
   taken for the first aggregate fissioning event. Let the cumulative
probability for the `first arrival' time for $D_i$ seeds be given by
   $F^{(D_i)}(t)$. Since each seed grows independently, this can be
related to the cumulative probability for the `first arrival' time
for a single seed via the relation
\begin{equation}
F^{(D_i)}(t) = 1 - (1 - F^{(1)}(t))^{D_i}
\end{equation}
The mean first arrival time $t_{1}$ is given by solving 
$F^{(D_i)}(t_1) = \frac{1}{2}$. For the simple probability distribution 
discussed above, the mean first arrival time is well-approximated by 
the expression
\begin{equation}
t_m(D_{i}) = t_0 + \frac{n_{1}}{2D_{i}} t_d
\end{equation}

2) We now proceed to calculate the time spent in the doubling phase,
{\it i.e.}, the time taken till the number of seeds reaches $D_f$. 
   All the aggregates formed after fissioning are assumed 
   to further fission in time $t_2$. Besides these, initial seeds
continue to `arrive', {\it i.e.}, aggregate to the fissioning size
and thus join the number seeds that are doubling. 
Let the number of such
seeds `arriving' in the time interval
   $m~t_2 < t  < (m+1)~t_2 $ be given by $\alpha(m) D$.
Then, $\alpha(m)$ is given by
\begin{equation}
\alpha(m) = F^{(1)}((m+1)t_2) - F^{(1)}(m t_2)
\end{equation}

3)  For our model distribution, the above equation can now be used to 
    calculate the mean incubation time, which is approximately given by 
\begin{equation}
t_i = t_0 + \frac{n_{1}}{2D_{i}} t_2 + [ {\mathrm log}_{2}(D_f) - {\mathrm log}_{2}(1+ \frac{D_i}{n_1})] t_2
\end{equation}

    The above equation gives the dose-dependence of the incubation time 
for the model distribution. It should be noted that this expression already 
explains several generic features seen in experimental dose-incubation 
curves (DICs) and in the microscopic models we present in later
sections. These are: 1) logarithmic dose-dependence at high doses
and 2) deviations from logarithmic behavior at low doses. From the above 
expression for the incubation time, it can be seen that the dominant 
contribution to the incubation time at high $D_i$ comes from the 
`doubling' phase which gives a logarithmic dose-dependence. However 
at low doses ($ D \le n1 $), the time spent in the doubling phase 
does not change appreciably with dose. Instead, the variation in 
the incubation time comes from the dose-dependence of the 
the `first arrival' time i.e. the lag phase time. This is reflected as 
a deviation from the logarithmic behavior in the DIC
which is seen in experimental DICs \cite{prusiner80}.
Our model thus makes the testable prediction that deviations from 
logarithmic behavior in the DIC should correspond to increases 
in the lag phase.

We now proceed to develop microscopic models for the initial 
aggregation process. The next section discusses the cellular automata 
approach to this problem.

\section{Cellular Automata Simulations}

Theoretical modeling of incubation times \cite{nowak98,eigen96,harper97}
starting at the molecular-level is all but impossible with a twenty
order of magnitude span between molecular motion time scales and those
of disease onset. On the other hand, kinetic theory allows one to
model long-time processes but ignores short distance spatial
fluctuations, important in nucleation and growth.
We have developed a 
lattice-based protein-level cellular-automata approach,
which bridges these two methodologies \cite{slepoy01}. 
Previously, we used it to calculate aggregation-time distributions, 
which compared favorably with the incubation-times
inferred from BSE data \cite{stekel96,anderson96}.
We also showed that playing with
the rules in such simple models can be a ``cheap'' way to suggest,
constrain and guide treatment protocols.

Our models consist of dilute concentrations of proteins diffusing on
the lattice and interconverting between their properly folded state
(PrP$^c$) and the misfolded state (PrP$^{Sc}$) \cite{cohen98}. 
Our key assumption is
that the conformational state of a protein depends on the amount of
water around it. A monomer isolated from others (surrounded by the
omnipresent water) stays in its properly folded state. However, when
proteins are surrounded by other proteins, thus excluding water from
parts of their neighborhood, they can change conformations and go into
a misfolded state (involving $\beta$ sheet bonding). A key parameter
of our model is the coordination, $q_c$, at which the misfolded
conformation PrP$^{Sc}$ becomes stable. Only misfolded monomers may
remain stably in a cluster, possibly breaking away from a cluster when
they fold back into the PrP$^c$ form.

Assuming aggregation happens on the cell-surface, we choose a 2d
hexagonal lattice. There is strong evidence that the local
coordination environment of the sphingolipids to which the prion
proteins attach is, in fact, hexagonal\cite{wille02}. The lattice
structure and the detailed protein motion are not crucial in our
model. At each time step, proteins can move randomly by at most a unit
lattice spacing. The magnitude of the time-step is set by the time for
a single monomer to misfold. It is implicitly assumed that proteins
co-adsorb with each other followed either by a conversion in shape or
separation. It is this conversion process which sets the unit of time.

By playing with the cellular automata rules it is possible to
get different aggregate morphologies and aggregation time distributions.
First, we consider the case where proteins are isotropic objects.
We have performed a large number of runs at values of $q_c=$ 1,
2, and 3, with different monomer concentrations (held fixed during the
simulation). The aggregation time is defined as the time required to
grow from initial seed of size ${\cal A}_i$ to a final size ${\cal A}$. 
The lower coordination rules effectively remove the nucleation
barrier, leading to frequent nucleation of new clusters.
Typical aggregate configurations (and
stable seeds) are shown in Fig. 1.

From these studies, we see that lowering the critical coordination
provides too rapid a growth for prion aggregates, and with no
nucleation barrier at these concentrations.  
In contrast, for $q_c=3$, the aggregation is very slow and it
is characterized by a broad incubation time
distribution, with a clear separation in time-scales for seeded and unseeded
({\it i.e.} infectious and sporadic) cases.

We can also obtain one dimensional fibril growth by considering the
proteins to be anisotropic. For example, on a square
lattice, we can get fibrils by: 
(i) identifying a preferred bonding face to our
simple point proteins, now made into squares. (ii) We choose a critical
coordination of 2. (iii) We make edge bonding of proteins with adjacent
preferred faces to be quite strong under coordination q=1 (i.e., the conversion
probability is higher than 50\%), and somewhat less strong for face to face
meeting of proteins. We assume zero conversion probabilities for all other
faces. By choosing three kinds of faces with appropriate rules, we can
obtain equivalent results on the hexagonal lattice. These rules assure
fibril growth which is dimer dominated (see Fig. 1).

\section{Stochastic Analysis of Aggregation}

In the low concentration limit, the aggregation results
from a sequential addition of proteins to the initial seed. However,
addition of monomers is not always stable. Given the rules,
various stages of the aggregate size and shape require a pair of
proteins (a dimer) to arrive simultaneously, in order to attach in a
stable manner. Thus, the entire process can be approximated by
one of stochastic sequential addition of monomer and dimer units. As
the concentration, $c$, goes to zero, the monomer addition rate is
proportional to $c$, whereas the dimer addition rate is proportional
to $c^2$, and thus the growth will involve a minimum number of dimers
and these will provide the dominant contribution to the growth times.

The growth to a final size ${\cal A}$ from an initial size ${\cal A}_{i}$ 
involves sequential addition of $n$ units.
The probability for the successive additions at intervals $t_1$,
$t_2$, $\ldots$, $t_n$ is
\begin{equation}
P(t_1,t_2,\ldots,t_n) = \prod_{j=1}^n p_{j} e^{-p_{j}t_j}, \label{eq:prod}
\end{equation}
where the rate for the $j$th unit, $p_j$, depends on the geometry of
the aggregate and the kind of unit (monomer or dimer) to be added.
Hence, the probability distribution associated with the total growth
time is
\begin{equation}
P(t) =\int_0^\infty dt_1\ldots\int_0^\infty dt_n
 \prod_{j=1}^n p_{j} e^{-p_{j}t} \delta(t-\sum_{i=1}^n t_i). \label{eq:pt}
\end{equation}
This integral can be evaluated by standard methods for arbitrary $p_i$.
We note the answers for two cases:

1) The attachment probabilities are identical for each unit i.e $p_j = p$
   for all $j$. In this case the probability distribution 
   is the Gamma distribution:
\begin{equation}
P(t) = \frac{p (pt)^{n-1}}{(n-1)!} e^{-pt} \label{eq:gamma}
\end{equation}
2) The rate, $p_j=p+j p'$, increases linearly with $j$. In this case we
obtain the Beta distribution in $e^{-t}$ \cite{szabo}:
\begin{equation}
P(t) = A e^{-(p+ p')t} (1 - e^{-p' t})^{n-1} \label{eq:pdim}
\end{equation}

In 1d fibril growth, dimers are attached one by one
with the area available for attachment of dimers staying constant, and
thus Eq. \ref{eq:gamma} applies. For 2d compact growth,
slow dimer attachments have to be combined with rapid filling up of rows by
monomers. In the low concentration limit, the rate is limited by 
dimer attachment probabilities which 
increase linearly with the number of dimers already attached, thus
leading to Eq. \ref{eq:pdim}.

At finite concentrations, the monomer attachment times can no longer
be neglected, and a more accurate treatment of the time scales in the
2d case requires a convolution of probabilities for
monomer attachment times with those for dimer attachment
times. The geometrical counting of number of monomers 
and number of dimers needed to grow to the desired size is
straightforward and can be used to develop accurate fits to the
numerical data (See Figure 2 A).

An important aspect of our 2d model is the asymptotic compression of
the distributions at low concentrations. The initial stage of the
growth is extremely slow and the process speeds up significantly as
the aggregate grows. Thus, the mean aggregation time, $t_m(1)$ to go
from an initial seed ${\cal A}_i$ to a final size ${\cal A}$ can be
much larger than the typical aggregate-doubling time $t_2$ to go from
size ${\cal A}/2$ to ${\cal A}$. Fig. 3 shows the ratio
$t_2(1)/t_m(1)$ for different concentrations and different final sizes
${\cal A}$. The crossover to monomer dominated behavior
($t_2/t_m(1) \approx \frac{1}{2}$) is indicated at the highest
concentration whereas at low concentrations this ratio can be much
smaller.

\section{Aggregate-Fissioning}

Fissioning of aggregates leads to exponential growth as the fission
products provide seeds for the next round of aggregation.  In this
subsection, we consider two mechanistic models of the fission process
associated with either proteolytic cleavage of aggregates or
mechanically induced breakage. We acknowledge that other models are
possible.

We assume that the fission time is small compared to the aggregation
time, and thus work in the limit of `instantaneous fission' in which
breakage of an aggregate happens much more rapidly than aggregation.
This implies a narrow distribution of fission sizes peaked, say,
at aggregate size ${\cal A}$, and in this limit our results are expected
to be independent of the width of the fission distribution.

We consider two extreme limiting models of fission: (i){\it Mechanical}.
In this case, once the aggregate reaches fission size ${\cal A}$, it
splits into two fragments of equal size ${\cal A}/2$. This
should approximately describe the situation in which aggregate size
is limited by nerve cell curvature (e.g., aggregation favors flat
planar
or linear structures, but the curvature of the neuron tends favors
curved structures). (ii) {\it Physiological}. In this case, the
aggregate can break into all smaller lengths at the fission scale.
This mimics the outcome of
protease attack for which there is no obvious preferred
site for breakage.

Taking a fixed background concentration of monomers, which
should be reasonable for at least short times in the disease.
The kinetic equation for the time evolution of aggregates with
size $n$ (measuring the number of dimers present)
and concentration $[a_n]$ is, for $n<{\cal A}$,
\begin{equation}
{d[a_n] \over dt} = p_{n-1}[a_{n-1}] - p_n[a_n]  + p_{f,n} [a_{{\cal A}}]
\label{eq:fission1}
\end{equation}
and, for $n={\cal A}$,
\begin{equation}
{d[a_{{\cal A}}] \over dt} = p_{{\cal A}-1}[a_{{\cal A}-1}] -
p_{f,0} [a_{{\cal A}}] ~~.
\label{eq:fission2}
\end{equation}
Here, for 1d aggregation $p_n = p_0$, while for the 2d aggregation
specified by our critical coordination three rules discussed in
Sec. III, $p_n=p+n p'$. 
For mechanical fission, $p_{f,n} = 2p_{f0}\delta_{n,{\cal A}/2}$,
while for the physiological fission, $p_{f,n} = 2p_{f0}/({\cal A}-1)$.
The instantaneous fission assumption requires $p_{f,0}>>p,p'$.

We can identify the doubling time from Eqns. \ref{eq:fission1} and 
\ref{eq:fission2}
by the following procedure: (i) Laplace transform the set of coupled
equations to obtain a matrix equation in transform space; (ii) identify
the largest positive eigenvalue of the Laplace matrix. In all
cases, we find but one positive eigenvalue. We have
systematically varied the fission size ${\cal A}$ and studied
the dependence of the exponential growth rate upon fission time.
For fibrils, the mean time to aggregate to size ${\cal A}$ is
$t_m \approx {\cal A}/p$, while for the 2d aggregates, the mean
aggregation time goes as
$t_m \approx \ln({\cal A})/p'$.
In the one dimensional case, we find that for large ${\cal A}$,
the doubling time $t_2$ tends to $0.5(0.43) t_m$ for mechanical(physiological)
fission. Hence, there is but a factor of two difference between
the aggregation time and the doubling time. Since the numerical
difference between mechanical and physiological fission is not substantial, 
we have examined only the mechanical
fission model for the 2d aggregate. In this case, we find that
the largest eigenvalue of the Laplace matrix goes as $\simeq 0.4/p$
independent of ${\cal A}$, while the aggregation time scales as
$\ln({\cal A})/p'$. Hence, for sufficiently large ${\cal A}$ it
is possible to make $t_2/t_m<<1$.

These results have the attractive feature of linking
the aggregation time, which we associate with the lag phase, to the doubling time
in fission.  However, we acknowledge that other processes may properly
describe fission. In particular, we cannot rule out continuous fissioning
of fragments off of large aggregates which may lead to a very different
result provided the fission rate is comparable to growth rates.

\section{Dose-Incubation curve}

In this section, we will look at the total incubation time and how it
varies as a function of the inoculated dose using the 
aggregation-time distributions
derived in the previous sections, and explore the extent to which
it provides a quantitative
description of experimental dose incubation curves.
As discussed before, a key advantage of the 2D growth
models is that, with a suitable fissioning scenario, they lead to
lag times much larger than doubling times. This is difficult to
accomplish with the 1D growth models. However, in this section
we will assume that the doubling time is an independent free parameter.
This allows us to fit the experimental dose incubation curves
by both 1D and 2D models. The constraints on relative values of
lag times and doubling times will be brought up in our discussions
in the next section.

We now proceed to calculate the incubation time as a function of
inoculated dose within our model for both 1d and 2d aggregation models
and compare with laboratory data. Consider first the DIC of the 263K
hamster scrapie strain. Kimberlin {\it et al} \cite{kimberlin86} have
determined the DIC along with independent measurements of the doubling
time $t_2$ and the final infectivity for this strain. The doubling
time $t_2$ can also be inferred from the DIC in the region where it
shows a logarithmic dose-dependence. Besides this experimental data,
we need to know the clearance ratio $r_i$ which gives the percentage
of the number of infectious seeds in a given inoculum which are
removed by rapid initial clearance. Let us illustrate this for the
case of 1 ${\mathrm LD_{50}}$ unit : in our model this corresponds to
having a 50$\%$ probability of attaching a {\it single} infectious seed.
Correspondingly the initial inoculum has to contain, on average,
$\frac{50}{100 - r_{i}}$ seeds (for $r_{i} > 50 \%$, typical values
from experiment are $r_{i} \sim 99\%$ \cite{manuelidis96}). We treat
$r_i$ as a parameter in our model to be determined by fitting the
DIC. Using this, in conjunction with our results for the aggregation
time distributions, we can generate theoretical DICs
using the method outlined in Section II.

For a fixed aggregation size ${\cal A}$, probability of attachment $p$
and infectivity ratio $r_i$, we then calculate the mean-square
deviation ($S^2$) (normalized by the experimental error estimates for 
each data point) between the theoretical and experimental DICs.
Minimizing $S^2$ gives us the optimal parameters, $p$ and ${\cal
A}$, for the particular strain for a given value of $r_i$ within our model.

We consider first the DIC for the 263K hamster scrapie strain. 
In carrying out the fitting, we have ignored the results at the highest 
doses since in this limit we are approaching the saturation of the 
incubation time.  
For our 2d growth model, we get a good fit to the experimental data as
indicated in Fig. 4a. The optimal parameters in this case are $ {\cal
A} = 16 $ (with ${\cal A}_{i} = 10$), $p = 0.025 ~day^{-1} $ and $r_i
= 0$ (which seems unphysical) for which $S^2 \sim 0.07$. 
For a more realistic value of $r_i \sim 88 \%$ we get 
$ {\cal A} = 16 $, $p = 0.16 ~day^{-1} $ with $S^2 \sim 1.14$.
From the fitting, one can see that the clearance ratio $r_i$
cannot be much greater than $r_i \sim 88 \%$ due to the constraints
imposed by the experimental incubation and doubling
times.  
It should be noted furthermore, that the above procedure does
not uniquely determine these parameters since comparably good fits are
obtained for higher values of ${\cal A}$ by correspondingly adjusting
$p$. For the 1-d growth model, we also get a good fit with $S^2
\sim 0.35$ for ${\cal A} = 8$ (with ${\cal A}_{i} = 4 $) and $p=0.11
~day^{-1}$.

One of the unusual features of the 263K scrapie strain in hamsters
is that at high doses the lag-time is negligibly small. This feature
is clearly seen in the time-course measurements \cite{kimberlin86} of
infectivity and also accords with the theoretical best-fit results
described above. In order to test our procedure for a more
representative strain we have also obtained a fit for the experimental
DIC for the ME7 strain in C57Bl mice \cite{taylor00}. The results
obtained by using our 2d growth model are shown in Fig. 4b. In
this case, the theoretical fit is not as good as that obtained for the
263K strain; the optimal parameters correspond to ${\cal A} = 140, p=
0.033 ~day^{-1}$ for $r_i = 99$ which gives $S^2 \sim 3.75$.
A comparably good fit was obtained by
using the 1d growth model with the optimal parameters ${\cal A} = 40,
p = 0.23 ~day^{-1}$ for $ r_i = 99$ which gave $S^2 \sim 5.02 $. However
in contrast with the 263K strain, the theoretical results for the ME7
strain give rise to a significant lag time of $ \sim 50 $ days
for the highest dose inoculated. This is a testable prediction
for time-course measurements of infectivity for the Me7 strain in
C57Bl mice.

Finally, we have used the above procedure to analyze the Sc237 scrapie 
strain in hamsters. In many respects, this strain is similar to 
the 263K scrapie strain in hamsters, however an analysis of the 
respective DICs reveals some significant differences. Based on the 
theoretical best-fit formula for the DIC \cite{prusiner99}, we can
estimate a doubling time $t_2$ for the Sc237 strain to be $\sim 2.1$ days  
whereas the doubling time for the 263K strain is $ \sim 3.9 $ days. 
This difference is also reflected in our theoretical best fit DIC 
curve for this strain which, for $ r_i = 99$, is given by the parameters 
$ {\cal A} = 80 $  $p = 0.052 ~day^{-1} $. For our 1d model the 
corresponding values are  $ {\cal A} = 62, p = 0.35 ~day^{-1} $.
Note that the experimental 
error estimates were not available for dose-incubation data for 
this strain, so we assumed them to 
be the same as that for the 263K strain in determining the best fit.
Using the above fits, we see that the theoretical prediction for the 
mean lag-time for an inoculated dose of 1 ${\mathrm LD_{50}}$ unit
is $\sim 50$ days. This value seems to be in good agreement with the 
experimental results for low dose inoculations for this strain 
\cite{prusiner99}.

The key results from our fitting (using the 2d growth model) are
summarized in Table I.  Based on the above results, we conclude that
while our model accounts for the features of the DIC and gives a good
fit to the experimental data, the latter cannot be used to distinguish
between the 1d and 2d growth morphologies or to ascertain the model
parameters conclusively.  However despite the ambiguity in the model
parameters, there are some robust predictions we can make after
fitting the DICs. We find that the experimental DIC, in
conjunction with our model, can be used to make predictions for the
time-course of infectivity: in particular we can predict the lag-time
as a function of dose. While the duration of the lag-time that we
calculate does depend on the clearance ratio $r_i$, we note that the
trend is that increasing $r_i$ reduces the lag-time. Thus by
measuring the lag-time at high doses we can determine the parameter
$r_i$ in our model, which then yields testable predictions for the
lag-time at low doses. In the case of the the Sc237 scrapie strain in
hamsters, our calculated lag-time at low dose agrees well with
experimental results for the same. For the other strains, the
predictions for the lag-time at low dose are a key testable prediction
of our model.

\section{Connection to Epidemiological Data}

We have found no dose incubation time data available for large mammals in the
literature, so to gain insight we have analyzed epidemiological data for
BSE in cattle\cite{anderson96}, a tabulation of incubation times from iatrogenic
CJD associated with {\it dura mater} transplants\cite{iatrogenic}, and the cluster of five victims of vCJD from the
village of Queniborough in the United Kingdom\cite{quen,wallstreet,Quenthreecasestudy}.
The goal is to produce approximate estimates of aggregate size and growth time
scales for comparison to the small mammal data, and, potentially, to guide
future epidemiological and public health studies.
     
In brief, our assumptions and methods are as follows: \\
1) {\it Model Distributions} 
We apply only the 2d model in the dilute dose limit (suitable for
digested prions); 
comparable quality fits can be obtained
from the 1d model, but the estimated $t_2/t_m$ ratio 
consistently and strongly violates our mechanistic model result from
Sec. V, while the bound for the 2d case is satisfied. \\
2) {\it Number of Doublings}.  For mice and hamsters, with 1g brains, 30
doublings to incubation is typical.  Given that the mean cattle brain is
500g, and the mean human brain 1500g, we take $n_2=40$ doublings from
infection to incubation for cattle and humans.  \\
3) {\it BSE fits}.  Ref. \cite{anderson96}
provides a candidate incubation time distribution which best fits the
epidemic time course, and yields a mean incubation time $<t>$=5 years and standard
deviation of 1.3 years.  The width fixes $p'$ uniquely for 
given $n,l=1+p'/p$, with $p'$ weakly dependent upon $n$.
We readily calculate $t_m(n,l,p')$$\approx$ $(\psi(n+l)-\psi(l))/p'$,
and we take the difference $<t>-t_m$ to be $n_2t_2$, the length of the
exponential growth phase.  With a minimum $t_2$=5 days typical for hamsters and mice
(see Table II), the maximum $n$ value can be be found at given $l$. For small
aggregates, $n=6$ is taken
to be a plausible minimum value which bounds $t_2$ above as 15 days.
The results for $l=2,10$ are summarized
in Table II.  Importantly, we 
find the maximum aggregate size at fission to be of order 80-100
monomers, approximately independent of $l$ for $2\le l\le 10$.  \\
4) {\it Fits to iatrogenic CJD data}.  The data of Ref. \cite{iatrogenic} sharply
constrain the size of the aggregate at fission, since the ratio of standard
deviation ($\simeq 3$ years) to mean ($\simeq 5.8$ years) rules out $n>4$ for
all $l=1+p'/p$ values.  To fit the data we perform the following procedure:
a) We produce a parameter free estimate of $t_m$ 
by multiplying $\sigma$ by the ratio of $t_m/\sigma$ for a given choice of $n,l$; this
subtracted from $<t>$ yields the estimated doubling phase period (and $t_2$, assuming
$n_2=40$ as above).  (b) Using the estimated value of $n_2t_2$, we produce
a one parameter fit in $p'$ to the cumulative distribution $F^{(1)}(t-n_2t_2)$ for the
given $n,l$ choice.  We find for $n=3$, we can consistently fit {\it for all
$l$}, while for $n=4$, comparable quality fits are obtained for $l=2,3$, but for $l\ge 4$
$n=4$ is no longer a viable choice.  The resulting {\it robust} estimates
of incubation parameters are $t_2=5.4\pm 0.4$ years, $\sigma = 3.2\pm 0.2$ years,
and $t_2=7\pm 1$ days.  The attachment rates $p'$ vary systematically with
$l$, but are at most 0.22$\pm 0.042$/year.  (Error bars reflect 95\% confidence
intervals.)\\
5) {\it Fits to Queniborough vCJD incubation time data}.  
For vCJD, we have extracted an
estimated onset time distribution from the Queniborough cluster,
taken as incubation times, from the available
scientific and journalistic literature\cite{quen,wallstreet,Quenthreecasestudy}.
We assume that this is a single dose event, in the low dose limit.  
We determine the estimated mean $<t>$ from the five data points to be Feb.,
1998.  Using non-linear leas squares analysis, 
we fit the cluster data to the cumulative beta distribution
$F^{(1)}(t-<t>+t_m(n,l,p'))$, placing points at half steps of probability increment. 
We also estimated the upper and lower bounds
to $p'$ corresponding to 95\% confidence, which yields a corresponding range
for $t_m, \sigma$.   Unlike the iatrogenic CJD and
BSE data, the overall incubation time is unknown, which prohibits an estimate
of $t_2$.  We thus take a reasonable estimate for the maximum doubling
time at $t_2^{max}\le 30$ days.  Using the minimum estimate of 9.0 years
for the incubation time from epidemiological studies\cite{ghani01}, 
we thus estimate a minimum lag time of
$t_m^{min}=t_{inc}^{min}-n_2t_2^{max}=5.7$ years. 
The minimum aggregate size at fission which can
exceed this $t_m^{min}$ value at the upper 95\% confidence limit is
$n=4,l=2$ (16 monomers).  We can take the upper aggregation size limit
from BSE as a reasonable bound ($n=20,l=2$ or $n=10,l=10$ giving 80 monomers). 
We summarize our results 
in Table III, for two values of $l$.  We stress that our results
here are at best a crude guide to expected model fits since: (1) there are only
five data points to fit to and several model parameters, (2) the infection is
likely to be characterized by slow {\it heterologous} protein attachment for
short times and more rapid {\it homologous} attachment for long times, while we
assume a single effective attachment probability, and (3) there is no guarantee
that the postulated single event was in the low dose limit.   \\

Given the simplicity of our model and the assumptions made for fitting, we shall
emphasize the most robust results of our analysis.  
Our results are summarized in Tables II (BSE), III (vCJD). We find several
robust features:(i) 
For BSE, iatrogenic CJD, 
and vCJD, we find that a wide range of $l$ produce comparable
fits, corresponding to a different starting
seed size.  We have thus quoted values from $l=2$ and $l=10$, the former representing
nearly minimal seed sizes, the latter maximal plausible ones.
(ii) In all three cases, the standard deviation of the incubation time distribution is large (1.2 years for BSE, vCJD, and 3.2 years for iatrogenic CJD),
supporting the assumption of small dose. (ii)  
The attachment rates are all comparable (BSE: 0.6/yr;
iatrogenic CJD: $\le$ 0.2/year; vCJD: $\le 0.66$/yr) and are   
significantly smaller than for the small mammal analysis of 
the previous section ($\ge$ 11/year from Table I). (iii) 
For BSE and CJD, $t_2$ is comparable to values for small
mammals (eg., $\le$15 days for BSE, $8$ days for iatrogenic CJD). 
For all strains considered here with the exception of 263K for hamsters, 
ratios of $t_2/t_m$
are small.  We shall discuss the large $t_2$ values for vCJD below.
(iv) Aggregate sizes
are comparable for large and small mammals ($\le 80-100$ monomers for
BSE, for iatrogenic CJD with $l=2,n=3$ we get 16 monomers, and for vCJD
apparently tens of monomers also), compared with 16-360 for the small
mammals of Table I.  Hence in each case, given a 2 nm characteristic
protein size, the length scale is of order tens of nanometers.  (If the
aggregating entities are in fact oligomers as suggested by 
Ref. \cite{wille02}, we only double the 
length scale.) 

Hence, our analysis strongly suggests interspecies variation in incubation
times is dominated by $t_m$, given 
given similar doubling times and aggregate sizes.
In our model, variation in $t_m$ is governed by $p'$, which then
suggests a significantly higher homeostatic monomer concentration
in small mammals compared to large ones, a point we discuss further below.
  
Our fitting of the Queniborough data
also suggests that long ($\ge 20$ years) incubation times for
vCJD are implausible.  The maximum $t_m$ value found (at the upper 95\% confidence
limit) for $n=l=10$ is 9.3 years; even with the extreme $t_2=30$ day value, this
gives a 12 year incubation time.  We note that the most probable (best fit) $p'$
values yield larger and implausible $t_2$ values for the minimum epidemiological
estimate of the incubation time.  To the extent our model is applicable, this
implies 9 years likely overestimates the incubation period.  In consequence,
in conjunction with epidemiological analyses, our work suggest that the number of
vCJD infections is likely to number in the hundreds.  
   
\section{Discussion}

Recall that our basic hypothesis is that incubation times are controlled by
prion-aggregation around infectious external seeds on the neuronal
surface. Furthermore, in our calculations, the distribution of aggregation
times arises from the stochastic growth process from seeds of a given 
initial size.
That only a narrow range of seed sizes is relavant here maybe
motivated by size-sensitivity of
(a) the blood-brain barrier (b) attachment
probability and (c) transportability of the seeds. The lag phase
corresponds to growth from initial seeds to a characteristic
fissioning dimension ${\cal A}$, after which one gets a multiplication
in seeding-centers and an exponential growth in infectivity.

That there is a long lag time despite external seeding by
intracerebral inoculation \cite{manuelidis96,kimberlin88}, and a
doubling time which is typically significantly shorter
\cite{manuelidis96}, both of which become sharply defined at high
doses, seems to be a general feature of the prion diseases. Our 2d
compact aggregate model, with the assumptions of the preceding
paragraph, explains these facts. In particular, 2d compact aggregation
generates a broad distribution of aggregation times for a single seed
with a well defined sharp onset time ($t_0$) and mean aggregation time
($t_m(1)$). With increasing number of seeds ($D_i$), the distribution
of times for the first seed to reach the typical fissioning size
${\cal A}$ will narrow. Correspondingly the lag time, determined by
the first fissioning event, will become sharply defined and
concentrate at the onset time, which only weakly depends upon $D_i$.

The doubling time ($t_2$) is defined by fissioning and subsequent
growth from size ${\cal A}/2$ to ${\cal A}$. Since growth from
different seeds is independent, self-averaging (`law of large
numbers') gives a sharply defined $t_2$. Thus at high doses both the
lag time and doubling time are sharply defined which accounts for one
of the most striking features of prion diseases: the reproducibility
of incubation times at high doses. Indeed, we can explain several
features of the dose-incubation curve. Notably, above a saturation
dose $D_s$, the incubation time does not decrease, while for $D<D_s$,
the incubation time varies as ${\rm log}(D)$, showing deviations from
the log only below a much smaller value $D_{min}$
\cite{prusiner82b,masel99}. The total incubation time is the sum of
the lag time and $n_dt_2$, where $n_d$ is the number of doubling
steps.  Assuming that the onset of clinical symptoms is related to the
damage of a fixed number of neurons \cite{nowak98}, the logarithmic
dose dependence of the number of doubling steps follows from the fact
that number of seeds grows exponentially in the fissioning stage. In
the range $ D_{min} < D < D_s$, the lag time does not change
appreciably with dose, thereby giving rise to the logarithmic
dose-dependence of the incubation time in this range. At low doses
($D< D_{min}$), the lag-time increases towards $t_{m}(1)$ (the mean
aggregation time for a single seed) giving rise to a broad
distribution of incubation times and the deviation from the
logarithmic behavior in the DIC which is observed
experimentally. \cite{prusiner80}

Furthermore, we note that in our mechanistic fissioning model,
the doubling time ($t_2$) is bounded 
above by the time to grow from size ${\cal A}/2$ to ${\cal A}$. 
If the fission produces jagged fragments, these can be effectively 
filled by monomers which will accelerate the subsequent growth process.
This is only possible for 2d compact aggregates and not for 
1d fibrils, for which the exposed ends will always be limited to dimer growth.
This possibility may account for the effective `1/c' dependence
in the incubation time observed for transgenic hamsters with multiple
copies of the prion gene \cite{prusiner90}, 
noting that for hamsters the doubling phase appears to dominate 
incubation \cite{kimberlin86}.

A key difference between the 1d and 2d morphology (shown in section
IV) is that, within our model, in case of the latter (i) the lag-time can
be an order of magnitude larger than the doubling time. If the total time in
doubling-steps becomes large compared to the lag time, the overall
distribution will be relatively narrow. Thus, only in the case of 2d
growth can one get (ii) a wide distribution for the overall incubation
time, with a width comparable to the mean. Thus assuming (i)-(ii) to
result entirely from the growth processes discussed here, strongly
points to a 2d (or 3d) morphology as controlling the incubation times.

The early growth morphology clearly deserves further experimental
attention \cite{horiuchi99,rochet00}. Typically, the {\it in vitro}
morphology of prion aggregates has been found to be fibrillar
\cite{zanetti99}. Frequently large fibrillar aggregates are also
observed post mortem in brain tissues. The morphology and size scale
for aggregates that cause neuronal death and infection is not
known. One could argue that the reason why {\it in vitro} aggregates
are not infectious is because they do not have the proper
morphology. We speculate that the attachment to lipid membranes could
make a vital difference to the aggregation process, which is missing
in {\it in vitro} experiments. It would be very interesting to carry
out the {\it in vitro} studies of prion aggregation in presence of
lipid membranes.

An important byproduct of our analysis is the ability to predict the
time course of infectivity from the DIC.  Provided we take the
doubling time ($t_d$) as an independent experimental parameter, such
predictions are very robust and do not rely on many details of the
aggregation-fissioning model, including initial growth morphologies.
Such predictions are particularly significant since experiments which
measure the time-course of infectivity, and hence determine the lag
phase, are extremely expensive and time consuming. This dose
dependence of the lag phase may well be a significant factor in
assessing the risk of infection.

Our results indicate that we can {\it infer} the (average) time-course
of infectivity using the information supplied by the experimental
DIC. Thus, based on our fits to the experimental DIC, we have made
testable predictions for the time-course of infectivity, and in
particular the lag time, as a function of dose. These predictions are
in good agreement with the existing experimental results and their
further experimental validation would prove very useful.

A factor which significantly affects the lag time is the probablity of
dimer attachment $p$; lowering $p$ increases the lag time. This is
relevant in understanding the species barrier effect in which there is
a reduction of incubation times with multiple passages in
inter-species infection \cite{lasmezas97}. During first passage, the
attachment of dimers is initially non-homologous but as the seed size
increases it should change to homologous attachment. Since the
non-homologous attachment probability should be smaller
\cite{horiuchi00}, the lag phase should be longer for first passage as
compared to subsequent passages. Thus, in our picture, most of the
difference in incubation times should come from the lag phase and the
exponential growth phase should be similar between first and second
passages. This has been observed experimentally for hamster scrapie
passaged in mice \cite{kimberlin79}.  We note that the estimated dimer
attachment rates for mice and hamsters range from 9-60 per year
(c.f. Table I), while for humans and cattle they are {\it maximally} 0.6 per
year.

Furthermore, we observe that the estimated aggregate sizes are comparable
between large animals (humans and cattle) and small animals, all in the
ballpark of tens of nm, which is precisely the estimated size of the lipid
rafts on which prions are hypothesized to rest\cite{rafts}.  The comparable
sizes of aggregates at fission we find in our model
between small and large mammals strongly suggests that the
primary determinant of lag time is the dimer attachment rate
which is regulated primarily by the
concentration (for a given strain).  This leads us to speculate
that the concentration of normal prion proteins must vary dramatically
between large and small species, which naturally leads us to envision
a link to metabolic rate.  Such a link is plausible if prion proteins
play a functional role in relieving oxidative stress as has been
proposed elsewhere\cite{oxidativestress}.  The hypothesis of enhanced
homeostatic PrP$^c$ concentration in small animals relative to large
ones is testable by direct examination of the brains of uninfected
animals.

Finally, as noted in the preceding section, our model analysis yields 12
years as an extreme upper bound on the vCJD incubation time.  We
actually get reasonable doubling times only when we assume a total incubation
below 9 years, which is the lowest epidemiological estimates.  The important
suggestion from this analysis is that the vCJD 
cumulative infection toll is likely
to be several hundred, vs. several hundred thousand\cite{ghani01}.  

{\it Acknowledgements}. We acknowledge useful discussions with
F. Cohen.  We thank D.D. Cox for a critical reading and discussion
of our fit to the Queniborough data.  R.V.K. and D.L.C. acknowledge
support from the U.S. Department of Energy, Office of Basic Energy
Sciences, Division of Materials Research. 
A.S. is supported by Sandia which is a multiprogram laboratory operated by 
Sandia Corporation,
a Lockheed Martin company, for the United States Department of Energy
under Contract No. DE-AC04-94AL85000. R.R.P.S. and D.L.C. have benefitted from
discussions at workshops of the Institute for Complex Adaptive
Matter. We are grateful for a grant of supercomputer time from the Lawrence
Livermore National Laboratory.

\begin{figure}[tb]
\epsfysize=6cm
\centerline{\epsffile{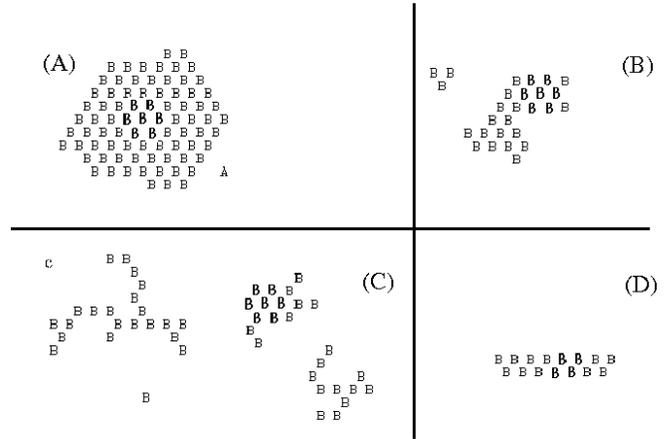}}
\caption{ Morphologies of seeds(bold Bs) and corresponding aggregates 
         due to the different rules: (A) $q_c=3$, (B) $q_c=2$,
         (C) $q_c=1$ and (D) fibril growth (see text) }
\label{fig1}
\end{figure}

\begin{figure}[tb]
\epsfysize=7cm
\centerline{\epsffile{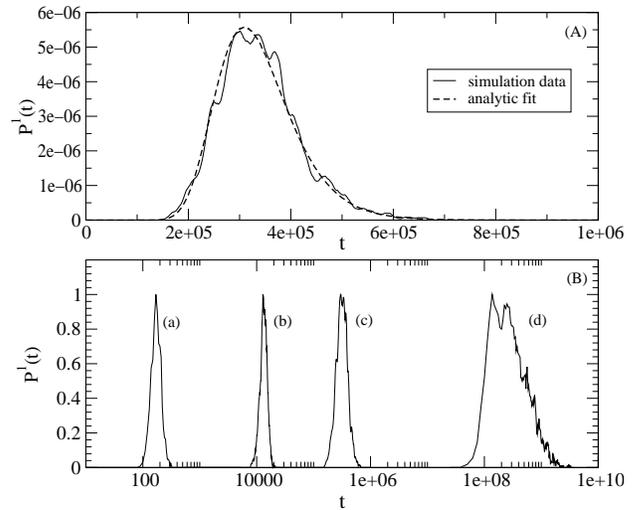}}
\caption{(A) Comparison of simulation data for single seed 
aggregation(${\cal A}_i $= 10, ${\cal A}$= 80, $c= 0.2\% $) 
and fit using analytical calculations (see text)
for 2d growth with q$_c$=3. The unit of time is 1 simulation sweep. 
(B) Probability distributions for (a) q$_c$ = 1, (b) q$_c$=2, 
(c) q$_c$=3 and (d) sporadic with q$_c$=3 at the same concentration
($c=0.2\%$). The maximum probability for
all distributions is scaled to unity. The sporadic result is obtained
by scaling the data at $c=1\% $ with an empirically determined $c^{-3}$ 
factor.} 
\label{fig2}
\end{figure}

\begin{figure}[tb]
\epsfysize=7cm
\centerline{\epsffile{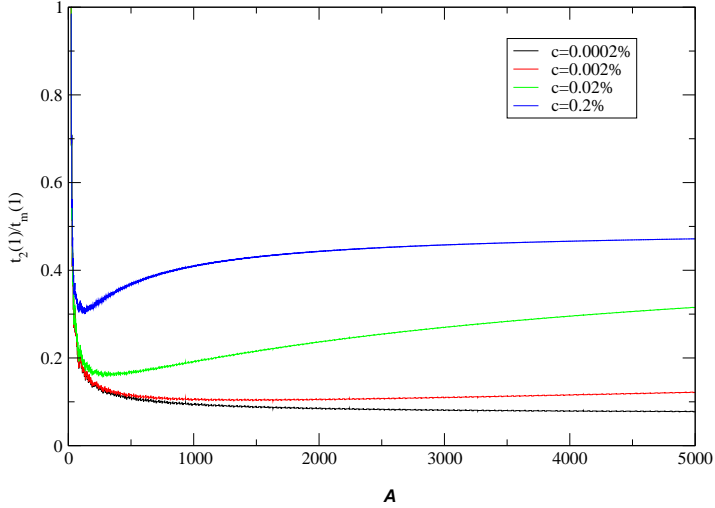}}
\caption{ Ratio of characteristic doubling time ($t_2$) to
mean incubation time ($t_m$) 
as a function of fissioning size ${\cal A}$  for single seed growth in 2d 
for different monomer concentrations, showing asymptotic compression 
as $c \rightarrow 0$. 
 }
\label{fig3}
\end{figure}

\begin{figure}[tb]
\epsfysize=7cm
\centerline{\epsffile{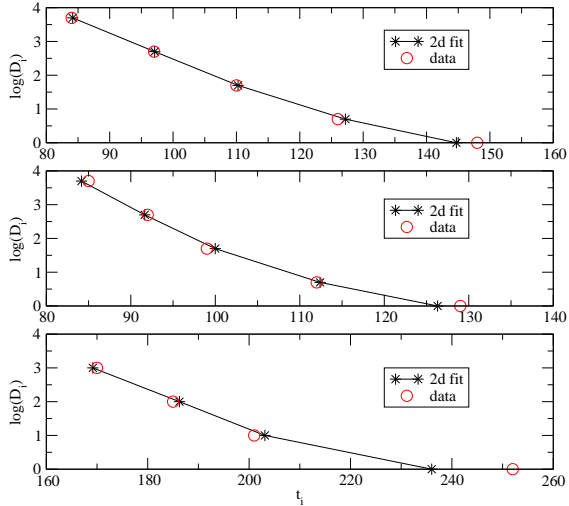}}
\caption{ (A) Experimental and theoretical dose-incubation curves for 
the 263K hamster scrapie strain for $r_i = 0$. 
The x-axis shows the incubation time and the 
y-axis shows the logarithm of the number of seeds inoculated. 
The theoretical curve is the best fit to the experimental data using the 
2d growth model for aggregation.
(B) Same as A but for the Sc237 strain in hamsters with $r_i =88$
(C) Same as A but for the Me7 strain in C57Bl mice with $r_i =88$
}
\label{fig4}
\end{figure}

\vskip 2cm

\begin{table}
\caption{Calculated best-fit parameters and predictions for low dose 
lag times for 3 scrapie strains.
$^a$ Ref.\ 28.
$^b$ Ref.\ 3. 
$^c$ Ref.\ 29. }
\begin{tabular}{|l|l|l|l|l|l|l|} \hline
Strain  &  $t_2$   & $r_i$   & Lag time    &    $p'$  &  ${\cal A}$ & $ S^2 $ \\
& (days)& & (days) & (days$^{-1}$)& &  \\ \hline \hline
 263K  &   3.9$^a$   &  0   &  27.7  & 0.025 & 16 & 0.07 \\ 
       &             &  88   &  4.3  & 0.16 & 16 & 1.14 \\ \hline
Sc237  & 2.1$^b$     & 88  & 56.9 & 0.052 & 140 & 0.11 \\ 
       &             & 99  & 48.9 & 0.052 & 80 & 0.11 \\ \hline 
Me7    & 4.5$^c$     & 88 &  107 & 0.033 & 360 & 3.86 \\ 
       &             & 99 & 89.6 & 0.033 & 140 & 3.75 \\ \hline 
\end{tabular}

\end{table}

\begin{table}
\caption{Fits to incubation time distribution for BSE (Refs.[11,33] ).  
}
\begin{tabular}{||l|l|l|l|l||}
$l$ & $n$ & $p'$  & $t_m$ & $t_2$ \\
& & (yrs$^{-1}$) & (yrs.) & (days)\\\hline\hline
2 & 6 & 0.57 & 2.8 & 20 \\
2 & 23 & 0.61 & 4.5 & 5 \\\hline
10 & 6 & 0.15 & 3.2 & 16 \\\hline\hline
\end{tabular}
\end{table}

\begin{table}
\caption{Fits to onset time distribution from Queniborough cluster (refs. 39-41). Here $l=1+p/p'$. Doubling times are computed for best fit $t_m$, values, and the upper and lower quotes for $p',t_m,\sigma$ represent 95\% confidence intervals. $t_2$ values are only quoted for best fit $p'$.   
 }
\begin{tabular}{||l|l|l|l|l|l||}\hline
$l$ &  $n $ & $p'$ & $t_m (t_m^{(+)},t_m^{(-)})$   & $\sigma(\sigma^{(-)},\sigma^{(+)})$   & $t_2$  \\
      &           &      (yrs$^{-1}$)    &  (yrs)     & (yrs) & (days)       \\\hline\hline
&&&&&\\
2 & 4    &  0.55(0.21,0.89)   & 2.3(1.4,5.4)  & 1.2(0.8,3.2)  &
61  \\
2 & 20   & 0.64(0.25,1.07) & 4.1(2.5,10.6) & 1.2(0.7,3.1) & 45\\
10 & 6   & 0.17(0.27,0.067) & 2.9(1.8,7.3) & 1.2(0.7,3.1) & 56 \\
10 & 10  & 0.20(0.077,0.32) & 3.6(2.2,9.3) & 1.2(0.7,3.1) & 37\\
\end{tabular}   
\end{table}


\begin{references}

\bibitem{prusiner82a} Prusiner, S.B., Gajdusek, D.C. and Alpers, M.P. 
{\it Ann. Neurol.} {\bf 12}, 1-9

\bibitem{prusiner82b} Prusiner, S.B., Cochran, S.P., Groth, D.F., Downey, D.E.,
Bowman, K.A. \& Martinez H.M. {\it Ann. Neurol.} {\bf 11}, 353-358

\bibitem{prusiner99}  Prusiner, S.B., Tremblay, P., Safar, J., Torchia, M. \&
DeArmond, S.J. (1999), in {\it Prion Biology and Diseases}, ed. S.B. Prusiner
(Cold Spring Harbor Laboratory Press, Cold Spring Harbor NY, 1999), p. 113-145

\bibitem{dickinson79}  Dickinson, A.G. \& Outram, G.W. (1979), in 
{\it Slow transmissible diseases of the nervous system}, 
ed. S.B. Prusiner \& W.J. Hadlow  (Academic Press, New York, 1979),
Vol. 2,  p. 13-32.

\bibitem{manuelidis96} Manuelidis, L. \&  Fritch, W. (1996) {\it Virology}
 {\bf 215}, 46-59

\bibitem{bolton98} Bolton, D. C. (1998) {\it J. Gen. Virol.} {\bf 79} 2557-2562

\bibitem{beekes96} Beekes, M., Baldauf. E. \& Diringer H. (1996)
{\it J. Gen. Vir.} {\bf 77}, 1925-1930

\bibitem{kimberlin88} Kimberlin, R.H. \& Walker C.A. (1988) in {\it Novel
infectious agents and the central nervous system} Wiley, Chichester (Ciba
Foundation Symposium 135), p. 37-62

\bibitem{mclean00} Mclean, A.R. \&  Bostock, C.J. {\it Phil. Trans. R. Soc.
Lond.} B (2000) {\bf 355}, 1043-1050

\bibitem{stekel96} Stekel, D.J., Nowak, M.A. \& Southwood, T.R.E. (1996) 
{\it Nature} (London) {\bf 381}, 119-119

\bibitem{anderson96}
Anderson, R.M., {\it et al.} (1996) {\it Nature}, {\bf 382}, 779-788.

\bibitem{kimberlin77} Kimberlin, R.H. \& Walker, C. A. (1977) 
{\it J. Gen. Virol.} {\bf 34}, 295-304

\bibitem{kimberlin79} Kimberlin, R.H. \& Walker C.A. (1978)
{\it J. Gen. Virol.}  {\bf 42}, 107-117


\bibitem{post00}
Post, K., Brown D.R., Groschup, M., Kretzschmar, H.A. \&
Riesner, D. (2000) {\it Arch. Virol.} (Suppl.) {\bf 16}, S265-S273

\bibitem{eigen96} Eigen, M. (1996) {\it Biophys. Chem.} {\bf 63},
A1-A18.

\bibitem{nowak98}
Nowak, M.A., Krakauer, D.C., Klug, A. \& May R.M. (1998) {\it Integr.
Biol.} {\bf 1}, 3-15.

\bibitem{masel99}
Masel, J., Jansen V.A.A. \& Nowak M.A. (1999) {\it Biophys. Chem.} {\bf 77}, 
139-152.

\bibitem{masel00} 
Masel J. \& Jansen V.A.A. (2000) {\it Biophys. Chem.} {\bf 88}, 47-59.

\bibitem{payne98} Payne R.J.H. \& Krakauer D.C. (1998) 
{\it Proc. Roy. Soc. Lond. B.} {\bf 265}, 2341-2346. 

\bibitem{stumpf00} Stumpf M.P.H. \& Krakauer D.C. 
(2000) {\it Proc. Nat. Acad. Sc. USA} {\bf 97}, 10573-10577. 

\bibitem{kellershohn01} Kellershohn N. \& Laurent M. (2001)
{\it Biophys. Jour.} {\bf 81} 2517-2529.

\bibitem{prusiner80} Prusiner, S.B., Groth, D.F., Cochran, S.P., Masiarz, F.R.,
McKinley, M.P., Martinez, H.M. (1980) {\it Biochemistry} {\bf 19}  4883-4891. 

\bibitem{harper97} Harper, J.D. \& Lansbury Jr., P.T. (1997)
{\it Ann. Rev. Biochem.} {\bf 66}, 385-407

\bibitem{slepoy01} Slepoy, A., Singh, R.R.P., Pazmandi, F., 
Kulkarni, R. V. \& Cox, D.L.
(2001) {\it Phys. Rev. Lett.} {\bf 87}, 058101

\bibitem{cohen98} Cohen, F.E. \& Prusiner, S.B. (1998) {\it Ann. Rev. Biochem.}
{\bf 67} 793-819

\bibitem{wille02} Wille, H., Michelitsch, M.D., Guenebaut, V., 
Supattapone, S., Serban A., Cohen, F.E., Agard, D.A. \& Prusiner, S.B.
(2002) {\it Proc. Nat. Acad. Sc. USA} {\bf 99}, 3563-3568. 

\bibitem{szabo} Szabo, A. (1988) {\it J. Mol. Biol.} {\bf 199}, 539-542

\bibitem{kimberlin86} Kimberlin, R.H. \& Walker C.A. (1986) 
{\it J. Gen. Virol.} {\bf 67}, 255-263. 

\bibitem{taylor00} Taylor, D. M., McConnell, I. \& Ferguson, C. E. (2000)
{\it Jour. of Virol. Meth.} {\bf 86} 35-40 

\bibitem{prusiner90} Prusiner, S.B., Scott M., Foster D., Pan K.-M., Groth D.,
Mirenda C.,Torchia M.,Yang S.-L., Serban D., Carlson G.A., Hoppe P.C.,
Westaway D., and DeArmond S.J. (1990) {\it Cell} {\bf 63}, 673-686

\bibitem{horiuchi99} Horiuchi, M. \& Caughey, B. (1999)
{\it Structure with Folding and Design} {\bf 7}, R231-R240.

\bibitem{ghani01} Ghani, A.C., Ferguson, N.M., Donnelly, C.M., Anderson,
R.M., (2001) {\it Nature}{\bf 406}, 583-584; Valleron, A.-J., Boelle, P.-Y.,
Will, R., Cesbron, J.-Y. (2001) {\it Science} {\bf 294},1726-1728; Huillard d'Aignaux,
J.N., Cousens, S.N., Smith, P.G. (2001) {\it Science} {\bf 294}, 1729-1731.

\bibitem{ferguson98} Ferguson, N.M., Donnelly, C.M., Woolhouse, M.E.J., Anderson,
R.M. (1998) {\it Phil. Trans. Ser. B R. Soc. London}{\bf 352}, 803-838. 

\bibitem{rochet00} Rochet, J.C. \& Lansbury P.T. (2000) 
{\it Curr. Op. Struc. Biol.} {\bf 10}, 60-68

\bibitem{zanetti99} Ionescu-Zanetti, C., Khurana, R., Gillespie, J.R.,
Petrick, J.S, Trabachino, L.C., Minert, L.J., Carter, S.A. \& Fink A.L. 
(1999) {\it Proc. Nat. Acad. Sc. USA} {\bf 96}, 13175-13179. 

\bibitem{lasmezas97} Lasm\'{e}zas, C.I., Deslys, J.-P., Robain, O.,
Jaegly, A., Beringue, V., Peyrin, J.-M., Fournier, J.-G., Hauw, J.-J., 
Rossier J. \& Dormont D. (1997) {\it  Science} {\bf 275}, 402-404.

\bibitem{horiuchi00} Horiuchi, M., Priola, S.A., Chabry, J. \&
 Caughey, B. (2000) {\it Proc. Natl. Acad. Sci. USA} {\bf 97}, 5836-5841

\bibitem{iatrogenic} Lang, C.J.G., Heckmann, J.G., Neund\"{o}rfer,
(1998) {\it J. Neurol. Sci.} {\bf 160} 128-139. 

\bibitem{quen}The five deaths were: August 1998, October 1998
(2), May 2000, October 2000, and from the scientific and journalistic
literature we have inferred that the observed onset dates were: August 1996,
September 1997, October 1997, and Jan 1999 (2).  There is a four month
uncertainty about one of the latter onset times. For a report, see\\
http://www.leics-ha.org.uk/cjd.htm and \\
http://www.rense.com/general4/cluster.htm.

\bibitem{wallstreet} Stecklow, S., (June 12, 2001) {\it Wall St. Journal} p. A1.  

\bibitem{Quenthreecasestudy} Allroggen, H., Dennis, G., Abbott, R.J., Pye, I.F.,
(2000) {\it J. Neurol. Neurosurg. Psychiatry} {\bf 68}, 375-378. 

\bibitem{rafts} Simons, K., Toomre, D. (2000) {\it Nature Reviews Mol. Cell Biol.}{\bf
1}, 31-39. 

\bibitem{oxidativestress} Guentchev, M., Siedlak, S.L., Jarius, C.,
Tagliavini, F., Castellani, R.J., Perry, G., Smith, M.A., Budka,
H. (2002) {\it Neurobiology of Disease} {\bf 9}, 275-281; Milhavet,
O., Lehmann, S. (2002), {\it Brain Res. Rev.}  {\bf 38}, 328-339;
Brown, D.R. (2001), {\it Trends. Neurosci.} {\bf 24}, 85-90.
\end{references}
\end{document}